\documentclass[fleqn,10pt]{wlscirep}

\title{DFT-inspired methods for quantum thermodynamics}

\author[1]{Marcela Herrera}
\author[1,2]{Roberto M. Serra}
\author[2,3]{Irene D'Amico}

\affil[1]{Centro de Ciências Naturais e Humanas, Universidade Federal do ABC, Avenida dos Estados 5001, 09210-580 Santo André, São Paulo, Brazil}
\affil[2]{Department of Physics, University of York, York YO10 5DD, United Kingdom}
\affil[3]{Instituto de Física de São Carlos, Universidade de São Paulo,  13560-970, São Carlos, São Paulo, Brazil}

\begin{abstract}
In the framework of quantum thermodynamics, we propose a method to quantitatively describe thermodynamic quantities for out-of-equilibrium interacting many-body systems. The method is articulated in various approximation protocols which allow to achieve increasing levels of accuracy, it is relatively simple to implement even for medium and large number of interactive particles, and uses tools and concepts from density functional theory.
We test the method on the driven Hubbard dimer at half filling, and compare exact and approximate results.
We show that the proposed method reproduces the average quantum work to high accuracy: for a very large region of parameter space (which cuts across all dynamical regimes) estimates are within 10\% of the exact results.

\end{abstract}
\begin{document}

\flushbottom
\maketitle
\thispagestyle{empty}

\section*{Introduction}

 The last decade has seen an increasing interest in the study of out-of-equilibrium thermodynamics at the micro- and nanoscale. 
 Such interest is impelled by the development of quantum technology and experimental control methods at small scales. In this scenario energy fluctuations play an important rule, thermodynamic quantities as work and entropy production are defined by their mean values, and the laws of thermodynamics still hold on average. The thermodynamic description of quantum many-body systems is significant for understanding the limits of the emerging quantum technology~\cite{Goold2016,Vinjanampathy2015,Millen2016,Parrondo2015,Liuzzo-Scorpo2016,Girolami2015}.  Fluctuation theorems~\cite{Jarzynski1997,Crooks1999,Esposito2009, Campisi2011,Jarzynski2011, Haenggi2015,Sagawa2012,Sagawa2013} have been key tools to describe the unavoidable fluctuations in the non-equilibrium dynamics and related experiments have been performed for small, non-interacting systems~\cite{Liphardt2002,Collin2005,Douarche2005,Toyabe2010,Saira2012,Batalhao2014,An2014,Batalhao2015,Auffeves2015,Koski2015,Peterson2016,Vidrighin2016,Rossnagel2016,Camati2016,Goold2016b,Cottet2017}, in both the classical and quantum domain.
When we turn to the case of systems composed of interacting particles ('many-body systems'), complex behavior arises as a consequence of the interactions. In general, dealing with a quantum many-body system entails a  tremendous effort and it is usually necessary to employ approximations to tackle the problem. Recently, some efforts have been directed to study the out-of-equilibrium thermodynamics in  many body systems like quantum harmonic oscillators' chains and spin chains \cite{Silva2008,Dorner2012,Joshi2013,Mascarenhas2014,Sindona2014,Fusco2014,Zhong2015,Eisert2015,Bayat2016,Solano2016}; however a method which can tackle a general system, and systems of increasing complexity is still lacking. In particular a Ramsey-like interferometry experimental protocol has been proposed in Refs. \cite{Dorner2013,Mazzola2013} to explore the statistics of energy fluctuations and work in quantum systems. This protocol has been so far implemented for single spin systems\cite{Batalhao2014,Batalhao2015} while its application to a quantum many-body system implies in fact considerable experimental (and theoretical) difficulties. This protocol requires not only experimental control of each individual part of the system and of the interactions between its components, but also the ability to construct conditional quantum operations where a single ancilla should be used as a control qubit for the collective, coherent behavior of the whole quantum many-body system. This is experimentally extremely challenging and even the theoretical simulation of the dynamics generated by the circuit would become prohibitive as the number of particles in the interacting system increases.

In this paper, we propose a new method to accurately describe some thermodynamics quantities (such as the mean work) in out-of-equilibrium quantum systems which can be applied to small, medium and large interacting many-body systems. In addition this method could be used to make the Ramsey-like interferometry experimental protocol accessible when consider many-body interacting systems.
Our method is inspired by density-functional theory (DFT) \cite{Jones1989,Dreizler1990,Argaman2000,Capelle2006,Jones2015} and will use some of the tools developed within it. DFT is one of the most efficient non-perturbative methods for studying the properties of interacting quantum systems \cite{Hohenberg1964}, and has produced tools widely used to describe diverse properties of many-body correlated systems,  such as band insulators, metals, semiconductors,  nanostructures, etc. \cite{Capelle2006,Jones2015}. At the core of DFT there is the mapping between the interacting many-body system onto a noninteracting one, the Kohn-Sham (KS) system\cite{Kohn1965}, which is characterised by the same ground state density of the interacting system.  In principle, from the density of such noninteracting system it is then possible to calculate exactly the ground-state (and even excited) properties of the interacting system, but in practice some approximations are required to obtain the so called exchange-correlation potential, a crucial quantity in DFT.  Despite, being  a very popular and successful approach to quantum many-body physics, to our knowledge, DFT or related methods have not been  applied, so far, to quantum thermodynamics. Here we will consider tools from a particular flavour of DFT, lattice-DFT\cite{Capelle2013}, but our method could be straightforwardly extended to different DFT flavours. Lattice-DFT applies to model Hamiltonians such as the Hubbard or Heisenberg Hamiltonians, which are particularly important for quantum information processing and that well describe the type of spin systems of interest to the quantum thermodynamic community.
 As a test-bed example we will study the out-of-equilibrium dynamics of the Hubbard dimer, which provides a variety of interesting regimes and behaviours to benchmark our method, since it can be solved exactly. We will compare exact and approximate results and show that our method provides very accurate estimates in most regimes of interest opening attractive further possibilities for applications.

\section*{The Hubbard Model}

The Hubbard model was conceived in 1963 separately by Gutzwiller \cite{Gutzwiller1963}, Kanamori \cite{Kanamori1963}, and Hubbard  \cite{Hubbard1963}  to describe the interaction of  electrons in solids.  This model gives  a microscopic understanding  of the  transition between  the  Mott-insulator and conductor systems \cite{Hubbard1964}, and allows for tunnelling of particles between adjacent sites of a lattice and interactions between particles occupying the same site. Here we consider the time-dependent one-dimensional Hubbard Hamiltonian \cite{Essler2016} described by 
\begin{eqnarray}
 \hat{H}(t)=-J\sum_{i=1,L;\sigma=\uparrow,\downarrow}(\hat{c}_{i,\sigma}^\dagger \hat{c}_{i+1,\sigma} +h.c.)+ \sum_{i=1,L}\Delta_i(t)\hat{n}_{i}+U\sum_{i=1,L} \hat{n}_{i,\uparrow}\hat{n}_{i,\downarrow},\label{Hubbard_Ham}
\end{eqnarray}
where $\hbar=1$ (atomic units), $L$ is the number of sites, $\hat{c}_{i,\sigma}^\dagger$ ($\hat{c}_{i,\sigma}$) are creation (annihilation) operators for fermions of spin $\sigma$,  $\hat{n}_{i,\sigma}= \hat{c}_{i,\sigma}^\dagger \hat{c}_{i,\sigma}$ is the site $i$ number operator for the $\sigma$ spin component, $\hat{n}_{i}=\hat{n}_{i,\uparrow}+\hat{n}_{i,\downarrow}$,  $J$ is the hopping parameter, $\Delta_i(t)$ is the time-dependent amplitude of the spin-independent external potential at site $i$, and $U$ is the on-site particle-particle interaction parameter.
Exact or numerically exact  solutions for static many-body problems are rare. For the Hubbard model, there exist an exact solution to the homogeneous one-dimensional case, however numerically exact solutions to the non-uniform case becomes quickly problematic as the number of particles and sites increases. Solutions to time-dependent many-body problems are even more difficult to achieve, and the Hubbard model is no exception. However quantum work due to an external driving of a many-body dynamics is one of those problems in which including time dependence and the effects of many-body interactions is essential to capture -- even qualitatively -- the system behavior.
In the following, we  propose a relatively simple approach to this problem which allows to include the key features stemming from the many-body interactions and dynamics, and yet maintaining the simplicity of simulating a non-interacting dynamics. We will show that, for a wide range of parameters, this simple approach allows to reproduce {\it quantitatively} the exact results.

\section*{DFT-inspired estimate of quantum thermodynamic quantities}

\subsection*{The Kohn-Sham system} The KS system is a fictitious, non-interacting, quantum system defined as having the same ground state particle density as the original interacting physical system \cite{Kohn1965}. The two systems have then the same number of particles. For each physical system, and a given many-body interaction, the KS system is uniquely defined \cite{Note1} and,  in the limit in which the physical system is non interacting, the Honehnberg-Kohn theorem  \cite{Hohenberg1964} ensures that the KS and the physical system coincide.

 Given an interacting system of Hamiltonian  $\hat{H}=\hat{K}+\hat{V}+\hat{V}_{ee}$, where $\hat{K}$ is the kinetic energy operator, $\hat{V}$ is the one-body external potential, and $\hat{V}_{ee}$ is the electron–electron repulsion, the corresponding KS system is described by the non-interacting Hamiltonian  $\hat{H}_{KS}=\hat{K}+\hat{V}_s$, where $\hat{V_s}$  is the one-body potential  $\hat{V}_s =  \hat{V}+ \hat{V}_H + \hat{V}_{xc}$. Here  $\hat{V}_H$ is the Hartree potential corresponding to the classical electrostatic interaction and $ \hat{V}_{xc}$  is the exchange-correlation potential, the functional derivative of the exchange-correlation energy $E_{xc}$, which contains additional contributions from the many-body interactions as well as the many-body contributions to the kinetic energy. Usually, the exchange-correlation potential is defined as a sum of exchange and correlation contributions, $V_{xc}=V_x+V_c$, and, due to its unknown functional dependence on the ground-state particle density, approximations have to be used for calculate it.

 \subsection*{Work in a quantum system} In a non-equilibrium driven quantum system work is defined as the mean value of a work probability distribution $\langle W\rangle=\int WP(W)dW$ that takes into account energy-level transitions (system histories) that can occur in a quantum dynamics \cite{Talkner2007}. In this scenario an external agent is performing (extracting) work on (from) a quantum system and the concept of work take into both the intrinsic nondeterministic nature of quantum mechanics and the effects of non-equilibrium fluctuations. The work probability distribution $P(W)$ contains all the information about the possible transitions in the Hamiltonian energy spectrum produced by an external potential (field) that drives the system out-of-equilibrium between the initial time $t=0$ and the final time $t=\tau$. This distribution is defined as $P(W)=\sum_{n,m}p_n p_{m|n} \delta\left(W-\Delta\epsilon_{m,n}\right)$, where $p_n$ is the probability to find the system in the eigenstate $|n\rangle$ of the initial Hamiltonian $\hat{H}(t=0)$,  and $p_{m|n} $ is the transition probability to evolve the system to the eigenstate $|m\rangle$ of the final Hamiltonian $\hat{H}(t=\tau)$ given the initial state $|n\rangle$.
 We will consider the work done on a system which starts in the thermal equilibrium state $\hat{\rho}_0=\exp\left[-\beta \hat{H}(t=0)\right]/Z_0$, where $\beta= 1/K_BT$ is the inverse temperature, $K_B$ is the Boltzmann constant, $T$ is the absolute temperature, and with $Z_t=\text{Tr}e^{-\beta \hat{H}_t}$ being the partition function for the instantaneous Hamiltonian $\hat{H}_t$ at time $t$. Then the system evolves up to time $\tau$ according to some driven protocol described by the time evolution operator generated by $\hat{H}(t)$ at the constant inverse temperature $\beta$. The final state of the system will not be, in general, an equilibrium state. This describes the non-equilibrium dynamics that we are interested in and it is the typical scenario explored in quantum thermodynamic protocols where fluctuations theorems can be applied.

\subsection*{DFT-inspired methods}
We consider the Hubbard system Eq. \eqref{Hubbard_Ham} and propose methods to accurately, quantitatively estimate the average quantum work $\langle W\rangle$ produced by its driven dynamics.
The key idea will be to  use the KS Hamiltonian $\hat{H}_{KS}$ as a zeroth-order Hamiltonian in a perturbation expansion scheme  which converges to the exact many-body Schrödinger equation\cite{Goerling1993,Goerling1994,Coe2008}. Using this expansion at its zeroth-order, gives a simple method of including interactions  within a {\em formally non-interacting} scheme through the DFT exchange-correlation ($\hat{V}_{XC}$) and Harthree ($\hat{V}_{H}$) terms. For the static case a similar scheme has been proven very effective to largely improve results over a wide range of parameters for the description of entanglement with respect to standard perturbation schemes\cite{Coe2008}.

The  formulation of DFT for the Hubbard model that we employ is the  site-occupation functional theory (SOFT) \cite{Gunnarsson1986,Lima2003}, where the traditional density in real space $n(\textbf{r})=\langle\sum_{\sigma}\hat{\psi}_{\sigma}^\dagger(\textbf{r}) \hat{\psi}_{\sigma}(\textbf{r})\rangle$ is replaced by the site occupation  $n_i=\langle\sum_{\sigma}\hat{c}_{i,\sigma}^\dagger \hat{c}_{i,\sigma}\rangle $.  Therefore, using  the SOFT we can write the KS Hamiltonian for the Hubbard model as
\begin{equation}
 \hat{H}_{KS}=J\sum_{i=1,L;\sigma=\uparrow,\downarrow}(\hat{c}_{i,\sigma}^\dagger \hat{c}_{i+1,\sigma} +h.c.)+\sum_{i=1,L}\Delta_{KS,i}(t)\hat{n}_{i}\label{H_0},
\end{equation}
where
\begin{equation}
\Delta_{KS,i}(t)=\Delta_{i}(t)+V_{H,i}+V_{xc,i}, \label{Delta_KS}
\end{equation}
and the Hartree potential is $V_{H,i}=U n_{i}/2$. The time-dependent potential $\Delta_{i}(t)$ defines the driven protocol and the exchange-correlation potential $V_{xc,i}$ reduces, for the Hubbard model, to the correlation potential $V_{c,i}
={\delta E_{c}}/{\delta n_{i}}$~\cite{Capelle2013}. However, as shown in the Methods section, not all the approximations for the exchange-correlation potential for the Hubbard model respect this property.
Using the KS Hamiltonian Eq.~\eqref{H_0},  we write the full many-body Hamiltonian as $   \hat{H}=\hat{H}_{KS}+\Delta\hat{H}$, where the perturbative term is then defined by
\begin{equation}
\Delta\hat{H}=-\sum_{i=1,L}\left({V}_{H,i} +V_{xc,i}\right) \hat{n}_{i}  +  U\sum_{i=1,L} \hat{n}_{i,\uparrow}\hat{n}_{i,\downarrow}.\label{H'}
\end{equation}

For obtaining  the mean value $\langle W\rangle$, we suggest three related protocols (as illustrated in Figure~\ref{scheme}). 

\subsubsection*{`Zero-order' approximation protocol}
We write the interacting Hamiltonian as $\hat{H}=\hat{H}_{0}+\Delta\hat{H}$ where $\hat{H}_{0}$ is a (formally) non-interacting Hamiltonian and $\Delta\hat{H}\equiv \hat{H}-\hat{H}_{0}$. We will consider the case in which $\hat{H}_{0}=\hat{H}_{KS}$ and compare it with the case in which $\hat{H}_{0}$ corresponds to the standard non-interacting approximation to $\hat{H}$, see Results section.
Then we approximate the initial thermal state of the system as the non-interacting one, $\hat{\rho}_0=\exp\left[-\beta \hat{H}_{0}(t=0)\right]/Z_0$, and the time-dependent evolution is calculated according to the (formally) non-interacting $H_0(t)$ up to the final time $\tau$. The quantum work will then be estimated based on this time evolution and on the spectra of the initial and final zero-order Hamiltonians. We note that in this protocol the time dependency in $H_0(t)$ is included only through the (explicitly) time-dependent term appearing in $\hat{H}$.
For $\hat{H}_{0}=\hat{H}_{KS}$, we expect this method to reduce the magnitude of the perturbation $\Delta\hat{H}$ as many-body interactions are already partially accounted for  the zero-order term $ \hat{H}_{KS}$, see Eqs. (\ref{H_0}), (\ref{Delta_KS}), and (\ref{H'}). We note that indeed  the zero order term $ \hat{H}_{KS}$ of this approximation already reproduces a very important property of the many-body system, namely the ground-state site occupation distribution. We therefore expect this to produce more accurate results with respect to the standard perturbation expansion.

\subsubsection*{Protocol with first-order correction to eigenenergies}
Here we propose to use the same formally non-interacting dynamics of the previous protocol, but now to include the first order correction in the estimate of the eigenenergies associated to the initial and final Hamiltonians of the system, with the corresponding correction to the approximation of the initial  thermal state.  As we will show, a better estimate of the eigenenergies may be  important to preserve agreement with exact results for certain regimes, and especially so when  the zero-order and exact Hamiltonians present qualitatively different eigenstate degeneracies, as in the case study below. While, in this paper, we will consider first-order corrections only, it should be possible to further improve accuracy by including higher order corrections to the eigenenergies.

\subsubsection*{Protocol including time-evolution effects of many-body interactions}
This third protocol applies to the case in which $\hat{H}_{0}=\hat{H}_{KS}$. Here we
include an implicit time-dependency in ${V}_{H}$ and ${V}_{xc}$. These quantities are functionals of the site occupation, so time dependence can be included via the time dependence of $n_i$. This implies a time-dependence which is local in time, i.e. that has no memory and could then not describe accurately non-Markovian processes. Nevertheless it allows to mimic, at least in part, the variation with time of the interaction effects due to the particles dynamics. As we will show, this significantly  improves results in certain regimes.  The time-dependent site occupation $n_i(t)$ will be obtained by solving the system self-consistently. This protocol takes inspiration from the adiabatic LDA approximation for time-dependent DFT  \cite{Runge1984,Verdozzi2008,Ullrich2012}. It may  be further enhanced by improving the approximation for the eigenergies of the initial and final time Hamiltonians, as described in the previous subsection.

We will consider two approximations for the KS exchange-correlation potential (see Results and Methods sections). We will compare the results from these protocols to the exact results and to  the results obtained for same-order standard perturbation theory.

 \begin{figure}
	\centering
  \includegraphics[width=11.5cm]{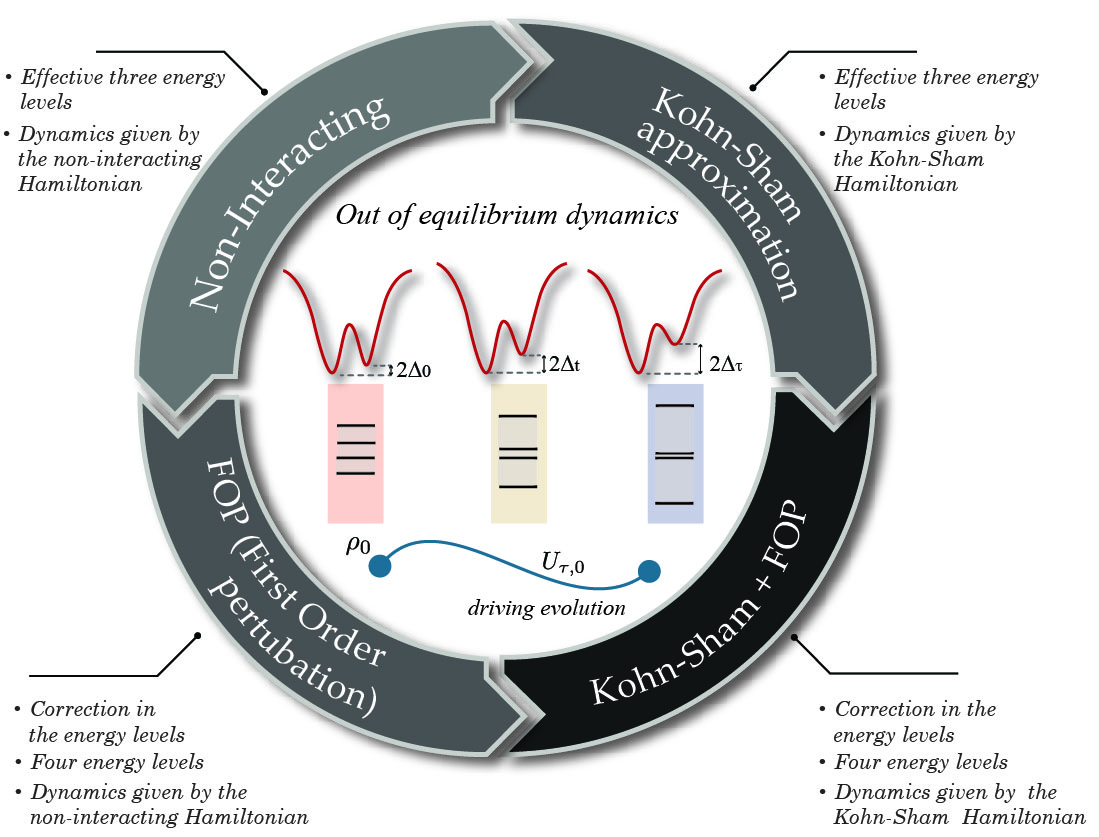}
  \caption{Illustration of the application of the protocols to the Hubbard dimer. This results in four possible approximations to obtain out-of-equilibrium dynamics and related thermodynamic quantities for the Hubbard dimer. These approximations can be used depending on the desired precision and the regime of interest. The worse-case scenario is to use a standard zero-order non-interacting Hamiltonian to describe the system and the dynamics. For more precision, and depending on the regime of interest, we can use either the Kohn-Sham Hamiltonian to describe system and dynamics, or use the standard zero-order non-interacting Hamiltonian for the dynamics but refine the approximation of the initial and final energy spectra (here done up to first order perturbation, FOP): the implementations of both options have similar degrees of difficulty. Finally, we can combine the Kohn-Sham Hamiltonian for the dynamics and the FOP (or higher order precision) for the initial and final spectra.}
  \label{scheme}
\end{figure}

\section*{Results}

We will focus on a Hubbard dimer with two electrons of opposite spin (half filling), which is known to display a rich physical behavior\cite{natphyseditoral2013,Murmann2015,Carrascal2015}, including  an analogue to the Mott metal-insulator transition \cite{Murmann2015,Carrascal2015}.  Because of its non-trivial dynamics, this model is ideal as a test bed for assessing the accuracy of approximations in reproducing quantities related to quantum fluctuations and quantum thermodynamics. When the system is reduced to a dimer with half-filling, the Eq.~(\ref{Hubbard_Ham}) can be written in the subspace basis set $\{|\uparrow\downarrow,0\rangle, |\uparrow,\downarrow\rangle, |\downarrow,\uparrow\rangle, |0,\uparrow\downarrow\rangle\}$  as
\begin{equation}\label{Hubbard_Ham_dimer}
 H=\left(
     \begin{array}{cccc}
       U+\Delta_1 & -J & J & 0 \\
       -J & 0 & 0 & -J \\
       J & 0 & 0 & J \\
       0 & -J & J & U+\Delta_2 \\
     \end{array}
   \right).
\end{equation}
We will calculate the average work along the dynamics driven by the linear time-dependent on-site potentials $\Delta_1=-\Delta_2=\Delta_0-(\Delta_0-\Delta_{\tau})t/\tau$, with the parametrized initial and final values $\Delta_0=0.5J$, $\Delta_{\tau}=5J$, at the parametrized temperature $T=J/(0.4 K_B)$.  The combined values of $U$ and $\tau$ will then determine the dynamical regime (sudden quench to intermediate to adiabatic, see discussion of Exact results). Due to the small Hilbert space associated to (\ref{Hubbard_Ham_dimer}), the quantum dynamics generated by it can be exactly solved by directly integrating the time-dependent Schr\"{o}dinger equation. The Hamiltonian (\ref{Hubbard_Ham_dimer}) can describe various physical systems, including, but not limited to, two gate-defined quantum dots \cite{Coe2010,Kikoin2012,Barthelemy2013}, cold atoms\cite{Murmann2015}, etc.

A schematic description of how we will apply to the Hubbard dimer the protocols proposed is provided in Figure~\ref{scheme}.

\subsection*{Exact results}

\subsubsection*{Average work}

\begin{figure}
	\centering
  \includegraphics[width=14.5cm]{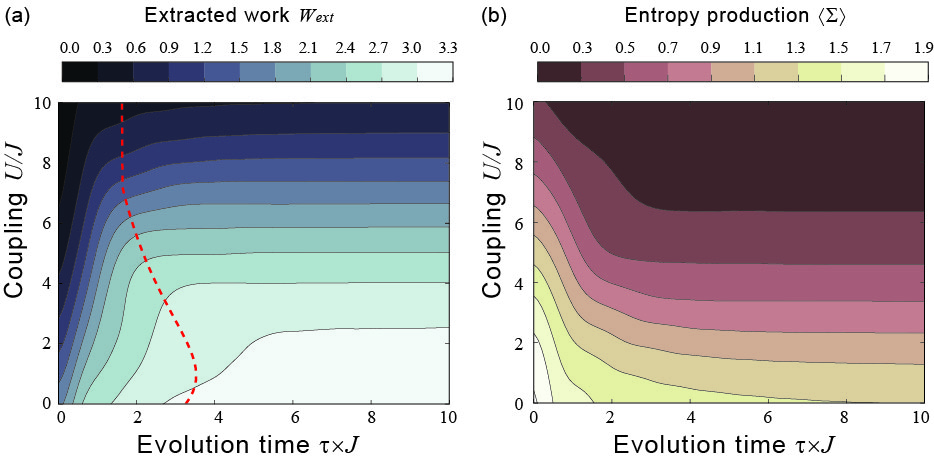}
  \caption{Mean extracted work, $-\left\langle W \right\rangle$, in units of $J$ and entropy production of a Hubbard dimer at half-filling. (a) Contour plot of the mean work and (b) contour plot of the entropy production for different values of the  particle-particle dimensionless interaction strength $U/J$ and the dimensionless evolution time $\tau\times J$. The red line in (a) marks  the change towards the adiabatic regime, when the transition probability between different Hamiltonian instantaneous eigenstates will become negligible along the dynamics. }
  \label{HD_exact}
\end{figure}

In the panel (a) of Figure~\ref{HD_exact} we display the exact work that can be extracted from an interacting Hubbard dimer at half-filling when operated according to the dynamics described by the Hamiltonian (\ref{Hubbard_Ham_dimer}). By varying the interaction $U$ and the time length of the dynamics $\tau$, we can access very different regimes: from non-interacting to very strongly interacting; from sudden quench, to intermediate and to adiabatic dynamics.

{\bf Crossover between non-adiabatic to adiabatic dynamics.} This system has three characteristic energy scales, $U$, $1/\tau$, and $J$. For the parameter region for which $U\stackrel{>}{\sim}J$, the crossover between non-adiabatic to adiabatic dynamics depends mostly on the interplay between $U$ and $1/\tau$, and occurs when the two energy scales become comparable, that is for $U\propto 1 /\tau$, see behavior of dashed-red curve in Figure~\ref{HD_exact}(a). For each given $U$, when  $\tau \gg1/U $ the work does not depend on the time length of the dynamics, showing that, for that particular $U$ the work has converged to its adiabatic value. For $\tau\stackrel{<}{\sim} 1/U $ and a given inter-particle interaction, the work instead strongly depends on $\tau$, increasing with $\tau$ up to the maximum allowed for that particular interparticle interaction. For $U\stackrel{<}{\sim}J$, the $J$ energy scale starts to influence the crossover between non-adiabatic to adiabatic regime, so that the simple relation between $U$ and $\tau$ described above dynamics breaks down in Figure~\ref{HD_exact}(a).

The contours curves for equal average work are strikingly different between the adiabatic and non-adiabatic regime: in the $U,\tau$ plane, they can be well approximated by $U=constant$ for the adiabatic regime, while they rapidly and almost linearly increase with $\tau$ for non-adiabatic dynamics. The behavior of these contour curves mirrors the fact that in the non-adiabatic regime the final state of the system, and so the work, is strongly influenced by the details of the dynamics, and hence strongly dependent on the time-scale on which the time-dependent driven protocol occurs; however, by definition, in an adiabatic (energy-level transitionless) process the system remains at all times with the same energy-level occupation of the instantaneous Hamiltonian as in the initial thermal state, which means that the final state of the system --and hence the work -- is completely defined, independently from the time taken by the system from going from the initial to the final state. We note that, as the Hamiltonian changes due to the driven protocol, the final state in the adiabatic regime is not an equilibrium state at the inverse temperature $\beta$.

{\bf Transition to `Mott insulator'.} The very strongly interacting parameter region  $U\stackrel{>}{\sim} 5J$ corresponds to the two-particle equivalent of the Mott-insulator~\cite{Murmann2015}, where double site occupation becomes energetically very costly. Hence for the Hubbard dimer, in this regime both double and zero site occupation are highly suppressed. As the Hilbert space available to the system dynamics reduces across the transition, we observe a corresponding decreasing of the average work that can be extracted from the system.

\subsubsection*{Entropy production and irreversibility}

The entropy production $\langle\Sigma\rangle$ is defined in terms of the dissipated work in the out-of-equilibrium dynamics, $\langle\Sigma\rangle= \beta\left(\langle W\rangle-\Delta F\right)$, where $\Delta F=-\left( 1/\beta\right)\ln\left( Z_{\tau}/Z_{0}\right)$ is the free energy variation in the protocol. It is related with an uncompensated heat, that is the energy that will be dissipated to the environment in order for the out-of-equilibrium system to get back to the thermal equilibrium. In this sense $\langle\Sigma\rangle$ quantify the degree of irreversibility of the dynamical process at hand~\cite{Batalhao2015,Velasco2011}. It is then instructive to look at it as the system undergoes through different dynamic regimes.
In general the degree of irreversibility will be related to the size of the Hilbert space available to the dynamics. For the system at hand {\it both} change to the adiabatic regime {\it and} Mott metal-insulator transition contribute to the reduction of the available Hilbert space, and hence to the decrease of entropy production.
This phenomenon shows well in the data plotted in the panel (b) of Figure~\ref{HD_exact}, where we observe a combined decrease in the entropy profuction as $\tau$ increases and the adiabatic regime is entered {\it and} as $U$ increases and the system tends to `freeze' towards the $n_1=n_2=1$ Mott-insulator configuration.

\subsection*{Results from `zero-order' approximations (standard non-interacting and Kohn-Sham based)}

In this section we compare results from the protocol which uses a {\it zero-order} approximation,  where the dynamics is propagated according to the Hamiltonian $\hat{H}_0(t)$ and time dependency is included {\it only} through the actual driving term $\Delta_i(t)$. As the {\it zero-order}  Hamiltonian, $\hat{H}_0 (t)$, is always {\it formally} the sum of single-particle Hamiltonians, the dynamics it generates will then correspond to the dynamics of non-interacting systems. Being formally non-interacting, this dynamics is easier to calculate numerically and would be easier to simulate and measure experimentally (in a quantum simulator) than the one originated by the many-body Hamiltonian $\hat{H}(t)$. We underline once more that the formally non-interacting systems in $\hat{H}_0 (t)$ represent physical systems (non-interacting particles) in the case of standard zero-order perturbation $\hat{H}_0 (t)=\hat{H}_{NI} (t)\equiv -J\sum_{i=1,L;\sigma=\uparrow,\downarrow}(\hat{c}_{i,\sigma}^\dagger \hat{c}_{i+1,\sigma} +h.c.)+ \sum_{i=1,L}\Delta_i(t)\hat{n}_{i}$, while represent fictitious systems (`Kohn-Sham particles') when $\hat{H}_0=\hat{H}_{KS}$.

\begin{figure}
  \centering
  \includegraphics[width=17.0cm]{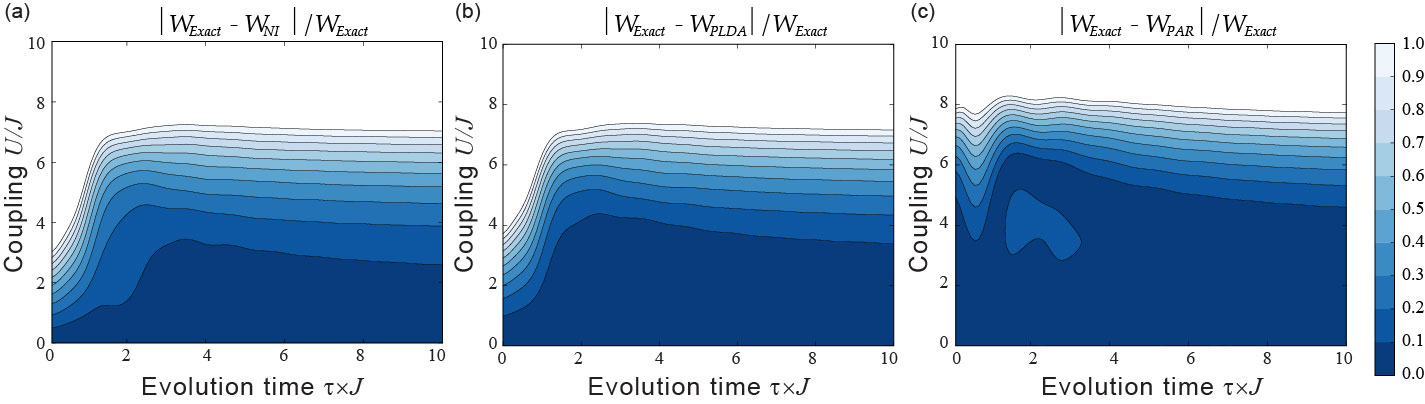}
  \caption{Relative error of the mean extracted work for the 'zero-order' perturbation protocol.  Contour plot of the relative error is shown as function of the dimentionless particle-particle interaction strength $U/J$ and the dimentionless evolution time $\tau\times J$. (a) Standard zero-order perturbation theory. (b) KS-based zero-order perturbation theory using the PLDA approximation for the exchange-correlation potential. (c) KS-based zero-order perturbation theory using an accurate parametrization for the exact $E_{c}$. }\label{LDA_no_FOP}
\end{figure}

In Figure~\ref{LDA_no_FOP} we present the contour plots of the relative error, with respect to the exact average extractable work for all `zero-order' approximations.
In the panel (a) we show the results for standard perturbation theory, where the dynamics is propagated according to the non-interacting part of the many-body Hamiltonian, $H_{NI}(t)$.
In the panels (b) and (c) of Figure~\ref{LDA_no_FOP} we present our results for the case where the Hamiltonian is approximated to zero-order by $\hat{H}_{KS}(t)$. We approximate the exchange-correlation potential in two ways: using the pseudo local density approximation (PLDA) as proposed by Gunnarsson and Sch\"onhammer\cite{Gunnarsson1986} (Fig.~\ref{LDA_no_FOP}, panel (b)) and using the recent parametrization to the exact  correlation energy $E_c$ proposed by Carrascal \textit{et al.}~\cite{Carrascal2015} (PAR)
(Figure~\ref{LDA_no_FOP}, panel (c)). The two approximations PLDA and PAR are described in the 'Methods' section.
It is known that the PLDA is not a particularly good approximation for the Hubbard model~\cite{Capelle2013}, but we chose it as we wish to show that even this provides already a good improvement over standard zero-order perturbation theory (compare panel (a) and (c) of Figure~\ref{LDA_no_FOP}). This is true especially for small and intermediate values of $\tau$: for example, for $\tau\approx 0$ (sudden quench), the use of PLDA instead of standard zero-order perturbation doubles the $U$-interval for which the relative error in the average work is below $10\%$, and makes it almost four times larger for $\tau\sim 2/J$ (non-adiabatic / adiabatic cross-over region).

The parametrization to $E_c$ by Carrascal \textit{et al.}~\cite{Carrascal2015} reproduces the exact correlation potential $V_{c}$ for the Hubbard dimer quite accurately; as such, Figure~\ref{LDA_no_FOP}, panel (c), may be taken to represent the best results we can expect for this system from the approximation with $\hat{H}_0=\hat{H}_{KS}$ we are proposing.
We see that now {\em the average work is reproduced to high accuracy in most of the $(U,\tau)$ parameter region}, even in parameter regions with strong many body interactions and/or corresponding to a dynamics very far from adiabaticity.

For very small $\tau$ (the sudden quench dynamics) we can reproduce the exactable work within 10\% accuracy for {\it interaction strengths larger than $U=4J$}. This is an enormous improvement over results obtained from the non-interacting zero-order dynamics (Figure~\ref{LDA_no_FOP}, panel (a)), where, for the same values of $\tau$, the exact exactable work could be reproduced within 10\% error only for $U\stackrel{<}{\sim} 0.5J$.
In the non-adiabatic / adiabatic cross-over region,  $\tau\approx 2/J$, we reproduce very well the exact average exactable work {\it up to interactions $U\approx 6J$}, while standard perturbation theory does poorly, accounting for interactions only up to $U\approx1J$ for $\tau\approx 2$ and up to $U\approx3J$ for $\tau\approx 3$. In this respect we wish to remark that even the `lighter patch' occurring in the panel (c) in Figure~\ref{LDA_no_FOP} within the region $3\stackrel{<}{\sim}U\stackrel{<}{\sim}5$, $1.5\stackrel{<}{\sim}\tau\stackrel{<}{\sim}3.5$ still corresponds to very good accuracy, with a maximum relative error of 12\% for the $\tau=2$ cut and of 14\% for the $U=4$ cut.
Finally, in the adiabatic regime, results from Carrascal's parametrization still outperforms substantially standard perturbation, almost doubling the 10\% relative error accuracy region, which now extends to interaction strengths of $U\approx 4.5J$, against the limit of $U\approx 2.5J$ for standard perturbation.

We note that, at least for the Hubbard dimer, a `zero-order'-type of approximation will always start to deteriorate as the system enters the Mott metal-insulator-type transition, and that this is independent of how well many-body interactions are accounted for in $\hat{H}_0$. In fact $\hat{H}_0$, and hence $\hat{H}_{KS}$, by definition, does not  include a many-body interaction term formally written as  $U=\sum_{i=1,L} \hat{n}_{i,\uparrow}\hat{n}_{i,\downarrow}$: this leads to a spectrum where the singlet and triplet eigenstates, $\frac{1}{2}[|\uparrow,\downarrow\rangle- |\downarrow,\uparrow\rangle]$ and  $\frac{1}{2}[|\uparrow,\downarrow\rangle + |\downarrow,\uparrow\rangle]$, are {\it always  non-degenerate}. However, in the Mott-insulator-type regime described by the actual many-body system of Hamiltonian $\hat{H}$, only the two aforementioned states remain energetically accessible and, most importantly,  {\it they become degenerate}. Why the first feature may be mimicked (e.g. this is done by the exact $\hat{H}_{KS}$ to reproduce the exact ground state site-occupation profile) the {\it intrinsic  qualitative difference} in degeneracy between the interacting and the formally non-interacting spectra determines the failure of any `zero-order'-type of approximation in the Mott-insulator-type region, which is what we observe in Figure~\ref{LDA_no_FOP}.

We note that the large improvement provided by the `zero-order', KS-based  approximations comes at no additional computational cost with respect to standard perturbation, as in both cases we are propagating formally non-interacting Hamiltonians.

\subsection*{Adding first order perturbation corrections to the initial and final energy spectra}

\begin{figure}
  \centering
  \includegraphics[width=17.cm]{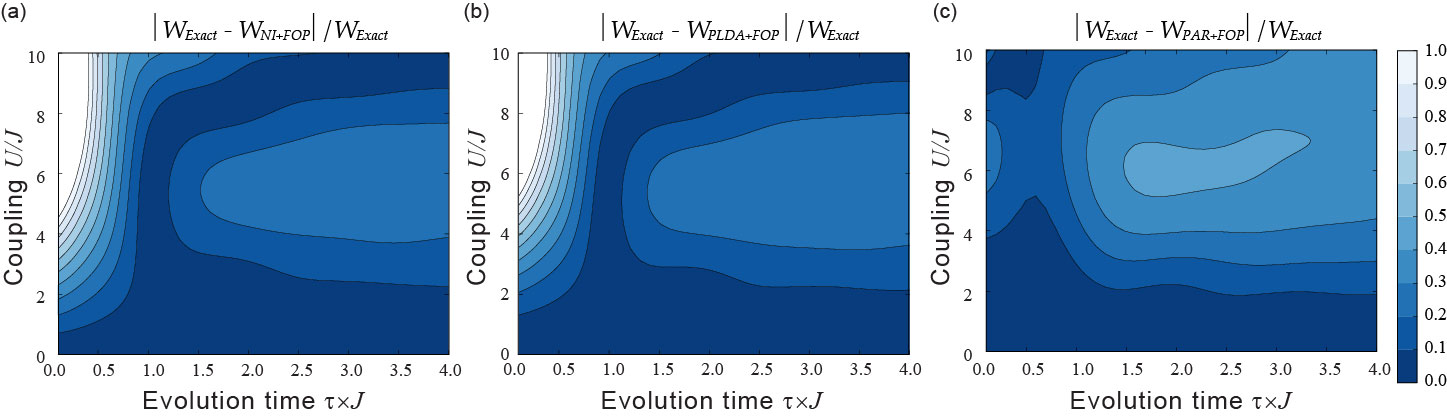}
  \caption{Relative error of the mean extractable work for the first-order correction to the eigenenergies' protocol.  Contour plot of the relative error is shown as function of the dimensionless particle-particle interaction strength $U/J$ and the dimensionless evolution time $\tau\times J$. (a) Dynamics is generated by standard zero-order perturbation theory. (b) Dynamics is generated by KS-based zero-order perturbation theory using the PLDA approximation for the exchange-correlation potential. (c) Dynamics is generated by KS-based zero-order perturbation theory using an  accurate parametrization for the exact correlation energy $E_{c}$. }\label{LDA_FOP}
\end{figure}

 At the end of the previous section we have discussed how, in regimes where the extent of extractable work is dominated by the spectrum and details of the system dynamics become less relevant, the KS-based `zero-order' approximation protocol may be seriously limited, and especially so if there exist different degeneracy patterns between the exact and the `zero-order' approximation spectra. For the Hubbard dimer this happens in the Mott-insulator-type parameter region.
In this section we explore if a potential solution to this issue could be to lift this degeneracy by applying higher order perturbation to the initial ($t=0$) and final ($t=\tau$) spectra, while performing the system dynamics according to the `zero-order' Hamiltonian.
In this paper, we have consider first order perturbation (FOP) corrections, and results are provided in Figure~\ref{LDA_FOP}. FOP corrections to the energy spectra can be considered accurate only for relatively low interactions $U\stackrel{<}{\sim}1J$.  For these values of $U$ we see indeed either an improvement in accuracy or, where results were already within the 10\% of the exact ones, this accuracy is maintained.

For larger values of the interaction, we can give a qualitative explanation of the influence of modifying the energy spectra. Let us first consider the adiabatic regime ($\tau \gg 1/U$): here results for the average work are dominated by the accuracy of the spectrum as the system -- in a perfectly adiabatic case -- would remain at any $t$ in a thermal state characterized by the same occupation probabilities determined at time $t=0$.
So for $2J\stackrel{<}{\sim}U\stackrel{<}{\sim}5J$, as the spectra provided by the FOP are increasingly {\it quantitatively} worsening, we see that this correction {\it reduces} the accuracy of the results. However, for $U\stackrel{>}{\sim}5J$ the system undergoes the analogue to the Mott metal-insulator transition, which, as discussed in the previous section, cannot be properly accounted for by 'zero-order' protocols because  there is a {\it qualitatively} different degeneracy between the zero-order and the exact spectra. In this parameter region then the protocol using the FOP spectrum, quantitatively inaccurate but qualitatively correct, provides a substantial improvement over the 'zero-order' protocol, as can be seen in Figure~\ref{LDA_FOP}.
In particular, FOP corrections are very important at very large particle-particle interaction strength $U$, $U\approx 10J$: here as long as $U$ is accounted for in the eigenenergy splitting, the system freezes in the ground state and the FOP approximation is then enough to reproduce the exact result $W=0$ shown in Figure~\ref{HD_exact}. In the same parameter region, without the FOP correction, the standard non-interacting zero-order approximation, completely independent from $U$, would predict maximum average work ($W\ge 3.3 J$ for $\tau \stackrel{>}{\sim} 2.5/J$), while the KS-based zero-order protocols provide some improvement over this result but still predict a way-too-high average work ($W\ge 2.1 J$).

For $U\stackrel{>}{\sim}1J$, and for non-adiabatic and transition regime  ($\tau \stackrel{<}{\sim} 1/U$), both spectrum and dynamics contribute to the average work. Here results from the FOP corrections seem to depend on how well the 'zero-order' protocol was already reproducing the exact dynamics. In particular, for the KS-based zero order protocol which uses the accurate parametrization of the exact $E_{c}$ (panel (c) of Figure \ref{LDA_FOP}), the contribution of the quantitatively incorrect spectra from the FOP worsen the results.

\subsection*{Including implicit time-dependency in many-body interaction terms(without and with FOP)}
So far we have considered zero-order Hamiltonians $\hat{H}_0$ where particle-particle interactions were included at most through time-independent functionals of the initial site-occupation. However a more accurate representation of the driven system evolution should be expected by including time-dependent functionals.
In this subsection we take inspiration from the adiabatic LDA and propose to include a time-dependence in these functionals by considering the same functional forms as for the static DFT but calculated at every time using the instantaneous site-occupation. The time-dependence considered is then local in time. To implement this protocol numerically, a self-consistent cycle to obtain the time-dependent
site-occupation $n_{i}(t)$, and from there the $V_{H,i}[n_{i}(t)]$ and $V_{xc,i}[n_{i}(t)]$ functionals, is necessary.

We illustrate this by applying the protocol to the PLDA exchange-correlation functional and focusing on the non-adiabatic and crossover regimes, $0\le\tau \le 4/J$.
We use as starting point the exact density at the initial
time, i.e., $n_{i}^{(0)}(t)=n_{i}^{(exact)}(0)$.   From this density
we obtain the exchange-correlation energy $E_{xc,i}^{(1)}(t)=E_{xc,i}^{(1)}[n_{i}^{(0)}(t)]$
and therefore the Kohn-Sham Hamiltonian $\hat{H}_{KS}^{(1)}(t)=\hat{H}_{KS}^{(1)}[n_{i}^{(0)}(t),t]$.
We evolve the system using this Hamiltonian and we obtain the
state of the system $\hat{\rho}^{(1)}(t)$. From this state we can calculate
the next iteration for the site-occupation  $n_{i}^{(1)}(t)=\text{Tr}\left[\hat{\rho}^{(1)}(t)\hat{n}_{i}\right]$.
Using this, we restart the same cycle calculating
the $E_{xc,i}^{(2)}(t)$ and consequently a new Kohn-Sham Hamiltonian
$\hat{H}_{KS}^{(2)}(t)$. This cycle is repeated until the convergence criteria  $\sum_{0<t<\tau}\lvert n_{i}^{(k-1)}(t)- n_{i}^{(k)}(t)\rvert/N = 10^{-6}$ is satisfied,  where the time $[0,\tau]$ is discretized in $N$ different values of $t$.

\begin{figure}
  \centering
  \includegraphics[width=14.cm]{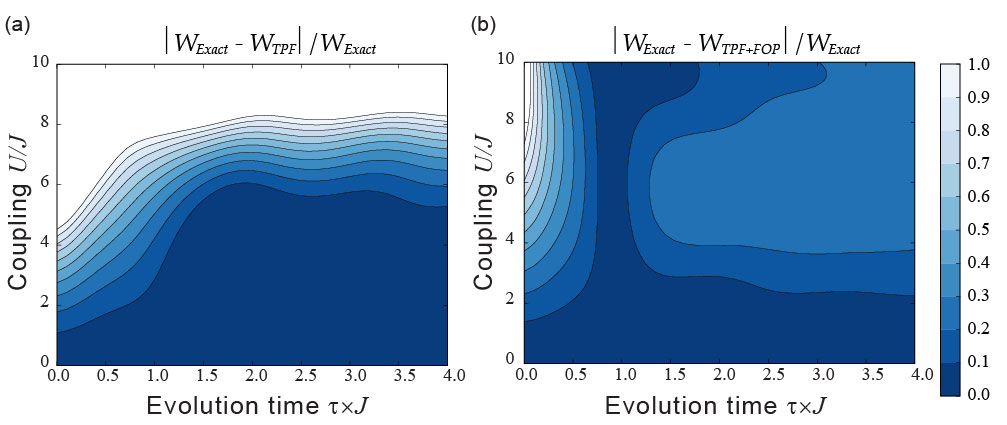}
  \caption{Contour plot of the relative error for the average extractable work when including time dependence of the functionals (TPF) describing many-body interactions. Data are presented with respect to the dimensionless coupling strength $U/J$ and the dimensionless evolution time $\tau\times J$. (a) panel: without FOP; (b) panel: with FOP.}\label{ALDA}
\end{figure}

Results are shown in Figure~\ref{ALDA}, panel (a), to be compared with the panel (c) of Figure~\ref{LDA_no_FOP} for $0\le\tau \le 4/J$. As the system exits the sudden-quench regime ($\tau\stackrel{>}{\sim}1/J$), and the site-occupation starts to respond to the dynamics, we notice a marked improvement over using time-independent functionals. For $\tau>1.5/J$ we now achieve an accuracy of at least 10\% up to interaction strength $5J\le U\le 6J$, while in Figure~\ref{LDA_no_FOP} it was only up to $U\le 4J$.
The wavy-pattern in the contour lines for $\tau>1.5/J$ reflects the system charge transfer dynamics between the two sites: the PLDA functional is unable to reproduce correctly the Mott-insulator transition so some charge transfer dynamics persists at large values of $U$.

When including first order corrections to the initial and final energy spectra (Figure~\ref{ALDA}, panel (b)), we recover, and for analogous reasons, a behavior similar to what observed in Figure \ref{LDA_FOP}.

\section*{Conclusion}

We have proposed a new method which uses some tools and concepts from density functional theory to study the non-equilibrium thermodynamics of driven quantum many-body systems, and illustrated it by the calculation of the average extractable work in a driven protocol. The method has the advantage of considering appropriate formally non-interacting systems (Kohn-Sham systems) to approximate the system dynamics, circumventing the theoretical and experimental problems of dealing with actual many-body interactions. It is easily scalable to large systems and can be used at different levels of sophistication, with increasing accuracy.

We have tested it on the Hubbard dimer, a two-spin system with a rich dynamics which includes the precursor to a quantum phase transition (Mott metal-insulator transition), and which can be embodied by various physical systems, including coupled quantum dots and cold atom lattices.
Our results show that the proposed method reproduces the average extractable work to high accuracy  for a very large region of parameter space:  for all dynamical regimes (from sudden quench, to the non-adiabatic to adiabatic crossover region, to the adiabatic regime) and up to quite strong particle-particle interactions ($U\stackrel{<}{\sim} 6 J$) our results are within 10\% of the exact results.

These very encouraging results, together with the simplicity of the method make for a breakthrough in the calculation of non-equilibrium thermodynamic quantities, as the quantum work, in a complex many-body system. Future developments include the possibility to combine the method with quantum simulation techniques and an experimental implementation of quantum simulators based on this method.

\section*{Methods}\label{methods}

\subsubsection*{Approximations for the exchange-correlation energy}
The exchange-correlation energy is a functional of the site occupation
density, but its functional form is unknown and needs approximations.
In this work we will consider and compare results from two different types of approximations
to the exchange-correlation energy for the Hubbard Model.

The first is the pseudo-LDA expression\cite{Gunnarsson1986}
\[
E_{xc}^{P-LDA}(n)=-2^{-4/3}U\sum_{i}n_{i}^{4/3}.
\]
Here the homogeneous reference system for the LDA is the three-dimensional electron gas, and so exchange is non-zero (for a related discussion see \cite{Capelle2013}).

In the second $E_{xc}=E_c$: this is the accurate parametrization to the exact correlation energy
recently proposed in \cite{Carrascal2015,Carrascal2015a}, and given by
\begin{equation}
E_{c}^{par}(\delta,u)=2J\left[f_{k}(\delta,u)-t_{s}(\delta)-e_{HX}(\delta,u)\right],\label{soft exc}
\end{equation}
where   $u=U/2J$, $\delta=\left|n_{1}-n_{2}\right|/2$, $f_{k}(\delta,u)=-g_{k}(\delta,u)+uh_{k}(\delta,u)$
, $t_{s}(\delta)=-\sqrt{1-\delta^{2}}$, and $e_{HX}(\delta,u)=\frac{u\left(1+\delta^{2}\right)}{2}$.

The function $g_{k}(\delta,u)$ can be obtained iteratively from the equation
\begin{equation}
g_{k}(\delta,u)=g_{k-1}(\delta,u)+\left[udh_{k-1}(\delta,u)-1\right]dg_{k-1}(\delta,u),
\end{equation}
and using as starting point
\begin{equation}
g_{0}(\delta,u)=\sqrt{\frac{\left(1-\delta\right)\left\{ 1+\delta\left[1+\left(1+\delta\right)^{3}ua_{1}(\delta,u)\right]\right\} }{\left[1+\left(1+\delta\right)^{3}ua_{2}(\delta,u)\right]}.}\label{g0}
\end{equation}
Here the coefficients $a_{1}(\delta,u)$ and $a_{2}(\delta,u)$ are given by
\begin{eqnarray}
a_{1}(\delta,u) & = & a_{11}(\delta)+ua_{12}(\delta),\\
a_{2}(\delta,u) & = & a_{21}(\delta)+ua_{22}(\delta),\nonumber
\end{eqnarray}
where $a_{21}(\delta)=\frac{1}{2}\sqrt{\frac{\left(1-\delta\right)\delta}{2}},$
$a_{11}(\delta)=a_{21}(\delta)\left(1+\frac{1}{\delta}\right)$,
$a_{12}(\delta)=\frac{1-\delta}{2}$, and $a_{22}(\delta)=\frac{a_{12}(\delta)}{2}$.

The functions $h_{k}(\delta,u),$ $dg_{k}(\delta,u)$, and $dh_{k}(\delta,u)$
are defined as
\begin{equation}
h_{k}(\delta,u)=\frac{g_{k}^{2}(\delta,u)\left[1-\sqrt{1-g_{k}^{2}(\delta,u)-\delta^{2}}\right]+2\delta^{2}}{2\left[g_{k}^{2}(\delta,u)+\delta^{2}\right]},
\end{equation}

\begin{equation}
dg_{k}(\delta,u)=\frac{\left(1-\delta\right)\left(1+\delta\right)^{3}u^{2}\left\{ a_{12}(\delta,u)\left[\frac{3\delta}{2}-1+\delta u\left(1+\delta\right)^{3}a_{2}(\delta,u)\right]-\delta a_{22}(\delta,u)\left[1+\left(1+\delta\right)^{3}ua_{1}(\delta,u)\right]\right\} }{2g_{k}(\delta,u)\left[1+\left(1+\delta\right)^{3}ua_{2}(\delta,u)\right]^{2}},
\end{equation}

\begin{equation}
dh_{k}(\delta,u)=\frac{g_{k}(\delta,u)\left\{ g_{k}^{4}(\delta,u)+3g_{k}^{2}(\delta,u)\delta^{2}+2\delta^{2}\left[\delta^{2}-1-\sqrt{1-g_{k}^{2}(\delta,u)-\delta^{2}}\right]\right\} }{2\left[g_{k}^{2}(\delta,u)+\delta^{2}\right]^{2}\sqrt{1-g_{k}^{2}(\delta,u)-\delta^{2}}}.
\end{equation}

In our calculations we used $g_{1}(\delta,u)$ to obtain the exchange correlation energy: this already provides good accuracy as shown in \cite{Carrascal2015}.

%\input{ref.bbl}
%\bibliography{ref}

\section*{Acknowledgements }

We thank D. J. Carrascal for valuable discussions. We acknowledge financial support from the University of York, UFABC, CNPq, FAPESP and from the Royal Society through the Newton Advanced Fellowship scheme (Grant no. NA140436). I.D. acknowledges support from CNPq (Grant: PVE—Processo: 401414/2014-0). This research was performed as part of the Brazilian National Institute of Science and Technology  for Quantum Information (INCT-IQ).

\section*{Author contributions statement}

I.D. and R.M.S. conceived the idea, M.H. performed the numerical simulations and contributed to refine the model. All authors discussed the results and contributed to the writing of the manuscript.

\section*{Additional information}

\textbf{Competing financial interests:}  The authors declare no competing financial interests.

%The corresponding author is responsible for submitting a \href{http://www.nature.com/srep/policies/index.html#competing}{competing financial interests statement} on behalf of all authors of the paper. This statement must be included in the submitted article file.
%Figures and tables can be referenced in LaTeX using the ref command, e.g. Figure \ref{fig:stream} and Table \ref{tab:example}.
\end{document}